\documentclass[sigconf]{acmart}
\usepackage{tikz}
\usepackage{pifont}

\newcommand{\ie}{\textit{i.e.}}
\newcommand{\eg}{\textit{e.g.}}
\newcommand{\yes}{\textcolor[rgb]{0,0.75,0}{\checkmark}}
\newcommand{\noo}{\textcolor{red}{$\times$}}
\newcommand{\mc}{\mathcal}

\AtBeginDocument{%
  \providecommand\BibTeX{{%
    \normalfont B\kern-0.5em{\scshape i\kern-0.25em b}\kern-0.8em\TeX}}}

\begin{document}
\fancyhead{}
%%
%% The "title" command has an optional parameter,
%% allowing the author to define a "short title" to be used in page headers.
\title[Pre-training for Information Retrieval: Are Hyperlinks Fully Explored?]{Pre-training for Information Retrieval: Are Hyperlinks Fully Explored?}

%%
%% The "author" command and its associated commands are used to define
%% the authors and their affiliations.
%% Of note is the shared affiliation of the first two authors, and the
%% "authornote" and "authornotemark" commands
%% used to denote shared contribution to the research.
\author{Jiawen Wu$^{1\dag}$, Xinyu Zhang$^{1\dag *}$, Yutao Zhu$^2$, Zheng Liu$^1$, Zikai Guo$^1$, Zhaoye Fei$^1$ \\ Ruofei Lai$^1$, Yongkang Wu$^1$, Zhao Cao$^1$, and Zhicheng Dou$^3$}
\affiliation{$^1$ Huawei Poisson Lab \country{China}}
\affiliation{$^2$ University of Montreal \city{Montreal} \state{Quebec} \country{Canada}}
\affiliation{$^3$ Gaoling School of Artificial Intelligence, Renmin University of China \state{Beijing} \country{China}}
\email{zhangxinyu35@huawei.com}
%%
%% By default, the full list of authors will be used in the page
%% headers. Often, this list is too long, and will overlap
%% other information printed in the page headers. This command allows
%% the author to define a more concise list
%% of authors' names for this purpose.
% \renewcommand{\shortauthors}{Trovato and Tobin, et al.}

%%
%% The abstract is a short summary of the work to be presented in the
%% article.
\begin{abstract}
Recent years have witnessed great progress on applying pre-trained language models, \textit{e.g.}, BERT, to information retrieval (IR) tasks. Hyperlinks, which are commonly used in Web pages, have been leveraged for designing pre-training objectives. For example, anchor texts of the hyperlinks have been used for simulating queries, thus constructing tremendous query-document pairs for pre-training. However, as a bridge across two web pages, the potential of hyperlinks has not been fully explored. In this work, we focus on modeling the relationship between two documents that are connected by hyperlinks and designing a new pre-training objective for ad-hoc retrieval. Specifically, we categorize the relationships between documents into four groups: no link, unidirectional link, symmetric link, and the 
most relevant symmetric link. By comparing two documents sampled from adjacent groups, the model can gradually improve its capability of capturing matching signals. We propose a progressive hyperlink predication ({PHP}) framework to explore the utilization of hyperlinks in pre-training. Experimental results on two large-scale ad-hoc retrieval datasets and six question-answering datasets demonstrate its superiority over existing pre-training methods. 
\end{abstract}
\maketitle

\def\thefootnote{\dag}\footnotetext{These authors contributed equally to this work.}
\def\thefootnote{*}\footnotetext{Corresponding author.}
\def\thefootnote{\arabic{footnote}}

\begin{figure}
    \centering
    \includegraphics[width=\linewidth]{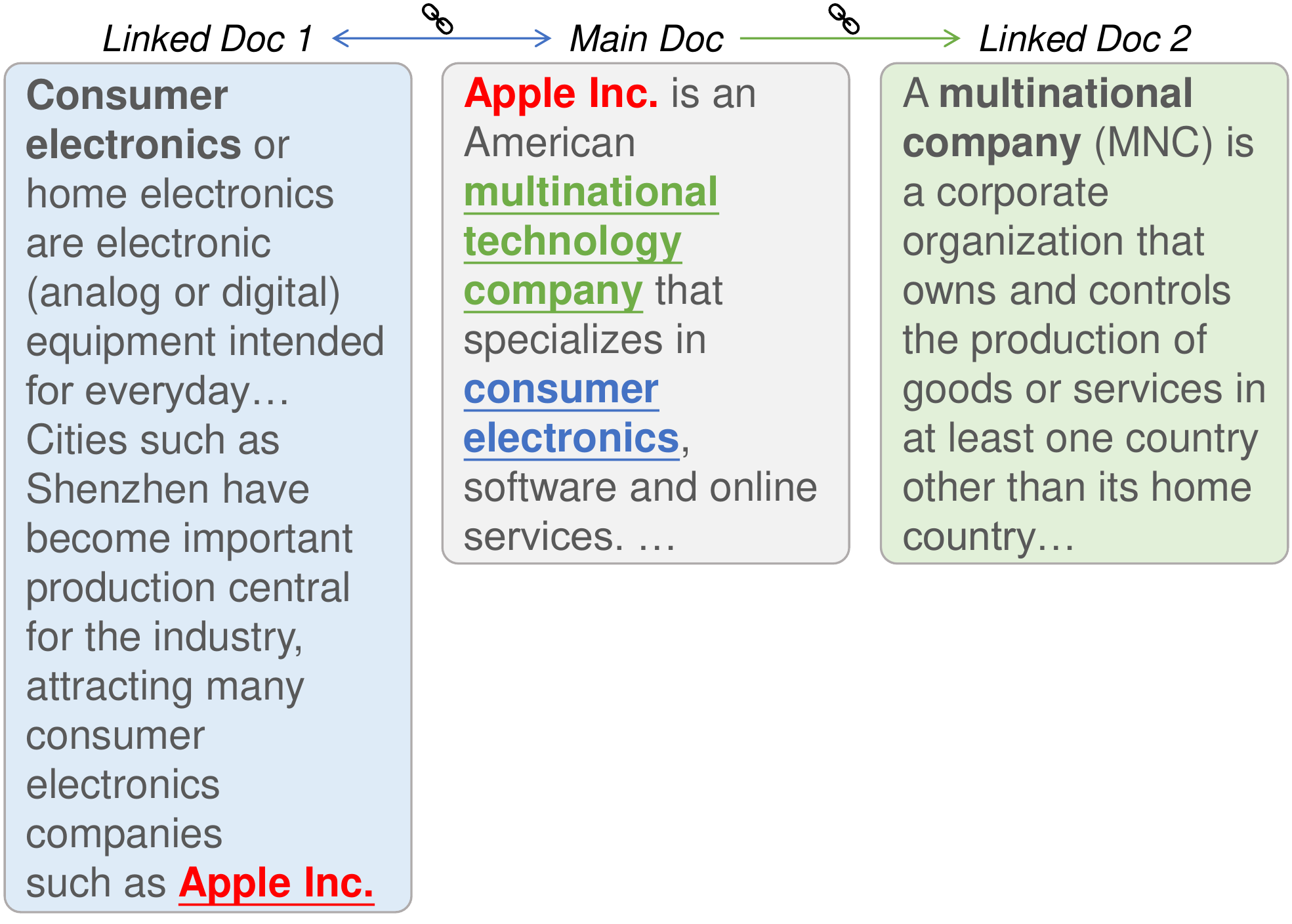}
    \caption{An example of linked documents. In addition to understand \textcolor{red}{Apple Inc.} from the document context, the two hyperlinked documents can provide some supplementary knowledge. The first linked document about \textcolor[rgb]{0.42,0.56,0.81}{consumer electronics} is more relevant than the second linked documents of \textcolor[rgb]{0,0.75,0}{multinational company} because it also mentions Apple Inc., constructing a back link to the original document.}
    \label{fig:exam}
\end{figure}

\section{Introduction}
% \begin{figure*}[t]
%     \centering
%     \includegraphics[width=\linewidth]{pic/example.pdf}
%     %   \vspace{-1.0em}
%     \caption{ The central document $d$ gives an introduction to Apple Inc. There are four documents around it displayed in the picture. As there exists an ``out'' hyperlink pointed to $d_1$ from $d$, $d_1$ is defined as an asymmetric linked document which describes the detail information of the term ``multinational''. $d_2$ and $d_3$ are symmetric linked documents to $d$ as there exists ``in'' and ``out'' hyperlinks. $d_2$ is the most relevant document according to our criterion introduced in Section~\ref{sec:MRDS}. Differently, there is no hyperlink between $d_4$ and $d$. Compared with $d_4$, all other documents $d_1$, $d_2$, and $d_3$ can provide some complementary information to $d$ by giving a detailed explanation of some term in $d$. Furthermore, compared with $d_2$ and $d_3$, $d_1$ shares less semantic correlation to $d$ as ``A multinational company'' is a more abstract concept to Apple Inc. Besides, when get the information of Apple Inc, ``it is a technology company'' is unique and most to know, while ``consumer electronic'' is just one of its specialization.}
%     \label{fig:example1}
%     % \vspace{-8px}
% \end{figure*}
Pre-trained language models have achieved great success in many natural language processing (NLP) tasks~\cite{DBLP:conf/acl-mrqa/DasGKGSYGGZZM19,DBLP:conf/iclr/AsaiHHSX20,DBLP:conf/naacl/DevlinCLT19}. They have also attracted numerous attention in the information retrieval (IR) area~\cite{DBLP:conf/cikm/MaDXZJCW21,DBLP:conf/wsdm/MaGZFJC21,DBLP:conf/iclr/ChangYCYK20,DBLP:conf/sigir/MaGZFLC21}. Many researchers have incorporated pre-trained language models, such as {BERT}~\cite{DBLP:conf/naacl/DevlinCLT19}, in IR tasks because of their powerful contextual representation capability. However, the conventional training objectives of the pre-trained language models in the NLP area are not optimal for IR problems. Basically, these pre-training objectives aim to make the model learn good contextual representations based on word co-occurrence information or the inter-sentence coherence relationship, whereas most IR tasks focus on modeling the ``relevance'' between a user-issued query $q$ and a document $d$ that provides the information to $q$. 
The problem goes beyond learning a good representation for individual documents or query sentences, but lies primarily in the relevance matching problem between queries and documents.

Many studies tackle this problem by designing different pre-training strategies to resemble the relevance between documents and queries~\cite{DBLP:conf/iclr/ChangYCYK20,DBLP:conf/wsdm/MaGZFJC21,DBLP:conf/cikm/MaDXZJCW21}. They hold the hypothesis that a well-designed pseudo query-document pairs in pre-training can benefit the downstream IR tasks. The objective is to make the query representation closer to the relevant document representation than the negative document representation in the embedding space.
For example, 
\citet{DBLP:conf/iclr/ChangYCYK20} proposed three paragraph-level strategies to construct the query-document pairs and conduct pre-training. \citet{DBLP:conf/wsdm/MaGZFJC21} sampled two words sets from a document as the query-document pair and \citet{DBLP:conf/cikm/MaDXZJCW21} leveraged hyperlinks and anchor texts to construct query-document pairs.

In summary, we can categorize exiting pre-training strategies into two groups: (1) The first group of methods focuses on learning language within a document~\cite{DBLP:conf/naacl/DevlinCLT19,DBLP:conf/iclr/ChangYCYK20}. Typical methods include masked language modeling (MLM) and next sentence prediction (NSP). They both simulate the human behavior of understanding a concept by its context. As shown in the middle part of Figure~\ref{fig:exam}, by masking \textit{Apple Inc.} and inferring it from the remaining sentence, we can understand Apple is a technology company selling consumer electronics. (2) The second group of approaches enhances the previous group by considering cross-document knowledge. This process is similar to the \textit{reference} behavior of humans. As can be seen in the example, when attempting to understand the concept of \textit{consumer electronics} in the main document, we can refer to another document (Linked Doc 1) for more sufficient information. This reference process is often achieved by hyperlinks in previous studies. 

% \dou{generally, I think the above motivation are not strange. They look TOO big and lack justification. For example, you claimed too many other applications, such as MRC and QA, but in this paper, you are just working on ranking}

% \dou{can you give some ``SIMPLE'' and important reasons? for example, some query-doc pretraining pairs are noisy and we need to refine them. }

% the multi-hop document retrieval task has proposed that a single piece of document cannot provide the complete information for a user-input query.
% Different from the conventional simple question answering task, the challenge of QA task has been shifted in the complex question which need to do inference over a document chain. The bottleneck of t
%  It has been proved that in most scenoria, the expected result is multiple pieces of document.

Existing pre-training methods have demonstrated the effectiveness of modeling cross-document knowledge through hyperlinks~\cite{DBLP:conf/cikm/MaDXZJCW21,DBLP:journals/corr/abs-2203-15827}. Typically, they assume that a hyperlinked document is more relevant than an unlinked one, and design the pre-training task by distinguishing whether two documents (segments) have a hyperlink. Unfortunately, the potential of hyperlink has not been fully explored. For example, as illustrated in Figure~\ref{fig:exam}, the main document hyperlinks to two documents. Their difference is that the first document also hyperlinks back to the main document, and from which we can know that \textit{Shenzhen is an important production central for Apple}. Intuitively, the first document is more relevant than the second one, and it indicates that documents with \textit{Shenzhen} can be ranked higher for answering the query \textit{Apple}. 

\begin{figure}[t]
    \centering
    \includegraphics[width=\linewidth]{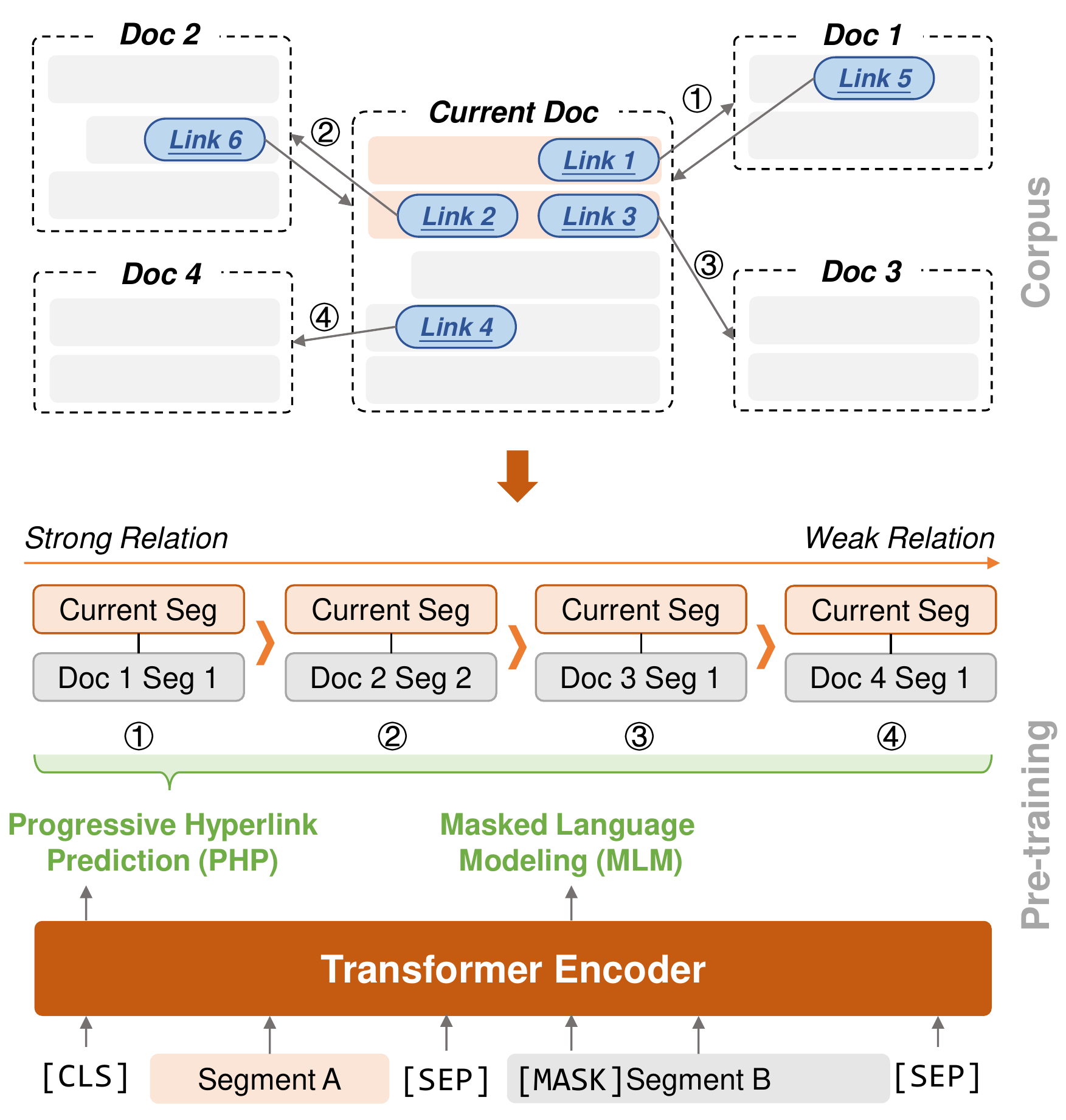}
    \caption{An overview of our {PHP} method. The Transformer encoder is pre-trained by our proposed progressive hyperlink prediction and masked language modeling task. After pre-training, the model will be fine-tuned on downstream tasks.}
    \label{fig:overview}
\end{figure}

In this work, we improve the previous works by introducing \textbf{P}rogressive \textbf{H}yperlink \textbf{P}rediction (PHP) objectives, where the language model is pre-trained to identify different hyperlink relationships following a curriculum learning-style workflow. Particularly, as shown in Figure~\ref{fig:overview}, we define the following four types of relationships for an anchor text segment and its linked neighbors: the strong symmetric linkage as the current doc and doc 1, 
% \emph{(wiki docs, for example, the segment of doc links doc 1, and the title of doc is at the front of doc 1)}
the weak symmetric linkage as the current doc and doc 2,
% \emph{(the segment of doc links doc 2, and the title of doc is at the back of doc 1)},
the strong asymmetric linkage as the current doc with doc 3,
% \emph{(the segment of doc links doc 3, and the title of doc 3 is in the segment)},
and the weak asymmetric linkage as the current doc with doc 4. 
% \emph{(the doc links doc 4, and the title of doc 4 is not in the segment)}. 
Detailed clarifications and definitions of them are given in Section~\ref{sec:method1}. 

\begin{figure}[t]
    \centering
    \includegraphics[width=.9\linewidth]{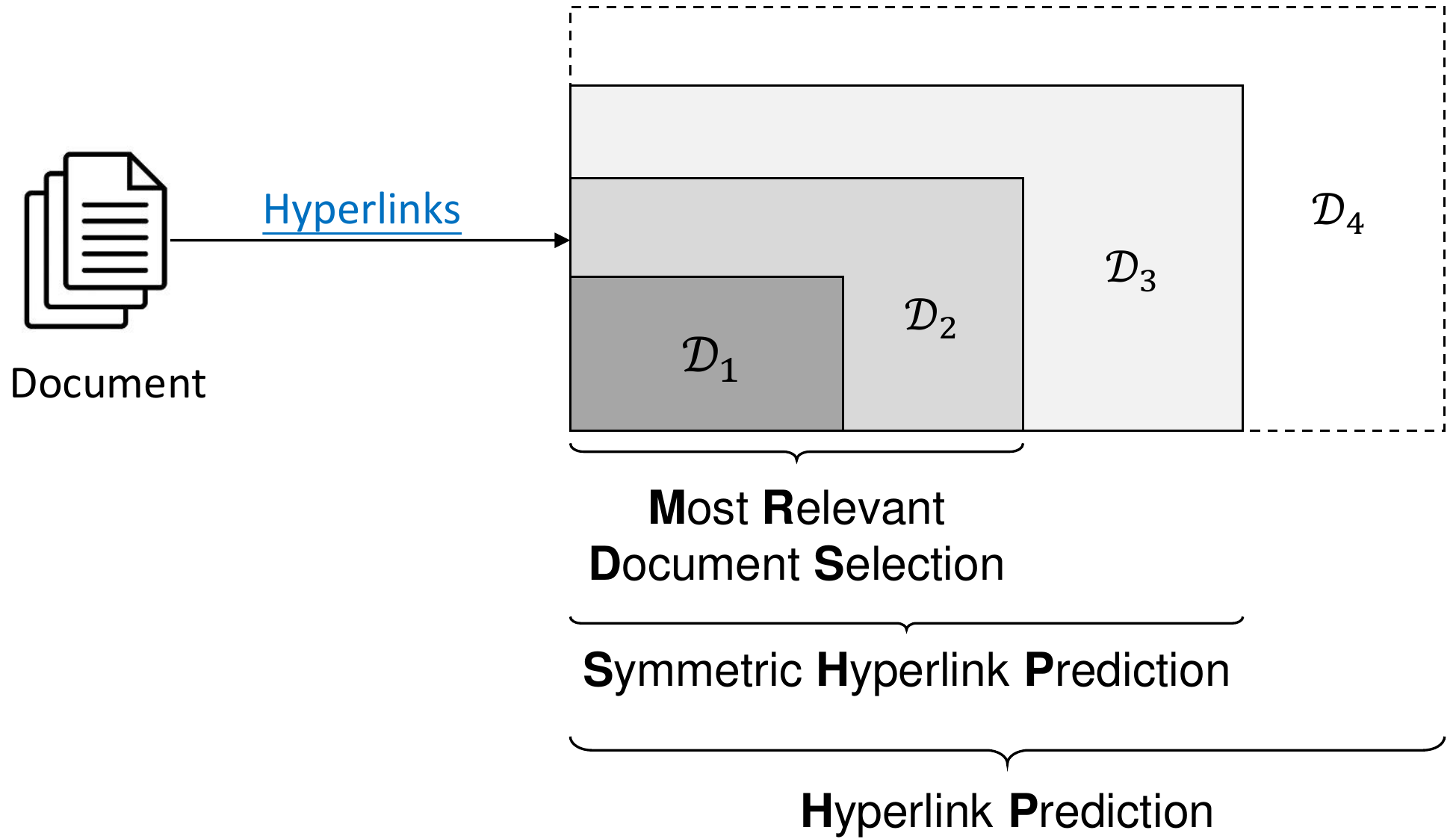}
    \caption{An illustration of four different groups of linked documents and three stages of pre-training. From outside to Inside, their relevance to the current segment is increasing.}
    \label{fig:document_levels}
\end{figure}

Different relationships above indicate different degree of semantic similarity. 
%\eg, the current doc is probably more semantically similar to doc 1 than the rest of documents given their higher hyperlink closeness. 
Essentially, we find that whether two linked documents mention each other and where they mention can represent a progressive relevance, as illustrated in Figure~\ref{fig:document_levels},  mention at the front part (strong symmetric linkage) > mention at the back pact (weak symmetric linkage), bidirectional mention (symmetric linkage) > unidirectional mention (asymmetric linkage).
In PHP, the model progressively develops its capability on semantic identification by learning to predict each of the relationships: it starts from the prediction of whether a hyperlink exists between two documents; then, it learns to predict whether the existing hyperlink is symmetric; and finally, whether the symmetrically linked documents establish a strong connection. Compared with the existing works learning the hyperlink relationships through a simple binary classification task, our proposed pre-training workflow turns out to be more challenging, hence forcing the language model to capture deeper semantic correlations between the documents.

We pre-train our method on Wikipedia articles with hyperlinks, and then fine-tune it on two ad-hoc retrieval datasets. Experimental results show that our method can achieve better performance than existing pre-training methods. In order to validates the generalizability of our method, we further fine-tune it on six extractive question answering datastes.
% and a natural language understanding benchmark (GLUE). 
The performance improvement over strong baselines clearly demonstrates the superiority of our method.

In summary, our contributions are three-fold:

% (1) We analyze the differences in hyperlinks at a finer granularity. The direction and position of the hyperlink are considered to identify documents' relevance. 

(1) We perform an in-depth analysis of hyperlinks and find that the link direction and position can indicate the different relevance of two documents. Consequently, we propose to categorize the hyperlinks into four groups. 
% We analyze the relevance indicated by different types of hyperlinks. 
% We propose a novel pre-training framework, where fine-grained information from the hyperlinks can be utilized for the enhancement of language model's capability on IR tasks. 

(2) We formulate three sentence-pair clarification tasks to identify the four groups of hyperlinks. These pre-training tasks are differentiated in terms of their difficulty, based on which we design a progressive learning workflow. By this means, the language model can effectively benefit from the pre-training process. 

(3) We conduct experiments on two ad-hoc retrieval daatsets. The evaluation results validate our advantage over the state-of-the-art pre-training methods. Further experiments on six question answering datasets clearly demonstrate the good generalizability of our method.

\section{Related Work}
\subsection{Pre-trained Language Models}
Pre-trained language models have been widely applied to various natural language processing (NLP) tasks and achieved outstanding performance. In general, these models are first \textit{pre-trained} on large-scale corpus, and then \textit{fine-tuned} on downstream datasets for specific tasks. By pre-training, the model can learn effective language representations, thus further improving the performance on downstream tasks. 
In the early stage, researchers explored pre-training Transformer~\cite{DBLP:conf/nips/VaswaniSPUJGKP17} on natural language corpus for language modelling. The great performance reveals the potential of pre-training. Typical works include {GPT} and {GPT-2}~\cite{2019gpt2}. 
% Based on the Transformer~\cite{DBLP:conf/nips/VaswaniSPUJGKP17} structure, researchers proposed to leverage large-scale corpus for pre-training. 
% researchers proposed some pre-trained generation model based on the powerful semantic modeling capability of Transformers, such as the GPT, GPT-2~\cite{2019gpt2}. 
Thereafter, researchers found that the standard conditional language models can only capture the information in a left-to-right or right-to-left manner, while the semantic information of the context is neglected, thus leading to sub-optimal performance on language understanding tasks. To tackle this problem, {BERT}~\cite{DBLP:conf/wsdm/DaiXC018}, \ie, the bidirectional transformer encoder, has been proposed. 
By training with the Masked Language Modelling (MLM) and Next Sentence Prediction (NSP) objectives, the model can better capture contextual information from both side, so its capability of modelling bidirectional semantics can be greatly enhanced. {BERT} has achieved the state-of-the-art performance on many natural language understanding tasks. Following {BERT}, many pre-trained language models are designed with various pre-training objectives~\cite{DBLP:journals/corr/abs-1907-11692,DBLP:conf/nips/YangDYCSL19,DBLP:conf/iclr/ClarkLLM20} or for different tasks~\cite{DBLP:conf/acl/LewisLGGMLSZ20,DBLP:journals/jmlr/RaffelSRLNMZLL20}.
Inspired by the powerful language modelling capability of {BERT}, the IR community has also developed some methods to better measure the relevance between queries and documents. A straightforward idea is treating the query and document as two sequences and fine-tuning {BERT}. Based on this, various works has achieved great performance on several IR tasks~\cite{DBLP:conf/sigir/DaiC19,DBLP:conf/ecir/GaoDC21,DBLP:journals/corr/abs-1901-04085,DBLP:journals/corr/abs-1910-14424,DBLP:journals/corr/abs-1904-07531}.
% Based on BERT, a lot of works have achieved good performance in some information retrieval tasks by enhancing pre-training data, and introducing more correlation information in the pre-training phase.

\subsection{Pre-training for IR}
Motivated by the great success of applying pre-training models to various areas~\cite{DBLP:conf/acl/TianGXLHWWW20,DBLP:conf/coling/ZhouTWWXH20}, many pre-training approaches are proposed specifically for IR tasks~\cite{DBLP:conf/iclr/ChangYCYK20, DBLP:conf/wsdm/MaGZFJC21,DBLP:conf/sigir/MaGZFLC21, DBLP:conf/cikm/MaDXZJCW21}. Existing methods have explored several methods on constructing well-pseudo query-document pairs from a variety of aspects during the pre-training stage, which simulates the down-stream query-document matching problem.
{PROP}~\cite{DBLP:conf/wsdm/MaGZFJC21} proposed a representative words prediction (ROP) task, which is inspired by the query likelihood model~\cite{DBLP:journals/arist/LiuC05,DBLP:journals/sigir/PonteC17}. The basic idea is that an optimal query should be generated as a piece of text representative of the ``ideal'' document. Given an input document, a pair of word sets are sampled according to a classical unigram language model, then the ROP objective is applied to predict the pairwise preference between them.
{B-PROP}~\cite{DBLP:conf/sigir/MaGZFLC21} further extended this method by introducing a powerful contextual language model, \ie, BERT,  to alleviate the limitations of the classical unigram language model. 
% A novel contrastive method is proposed to incorporate BERT. 
HARP \cite{DBLP:conf/cikm/MaDXZJCW21} declared the importance of the inherent structure of web data, \ie, hyperlink, in the retrieval phase. Specifically, they leveraged the correlations and supervised signals brought by hyperlinks and anchor texts, and devised four pre-training objectives to learn the semantic correlation between the generated pseudo queries and documents. 
\citet{DBLP:conf/iclr/ChangYCYK20} proposed three pre-training tasks that emphasise different aspects of semantics between queries and documents. They proposed three methods to construct pseudo query-document $q$-$d$ pairs, namely Inverse Cloze Task (ICT), Body First Selection (BFS), and Wiki Link Prediction (WLP). ICT takes a sentence sampled from a passage as a query $q$ and the rest of sentences as a document $d$; BFS takes a random sentence in the first section of a Wikipedia page as a query $q$ and a random passage in the same page as a document $d$; and WLP chooses the first sentence of a document $d_0$ as the query $q$, and a document $d_1$ hyperlinked  to $d_0$ as the document $d$. 

All existing approaches focus on constructing better pseudo-query document pairs in the pre-training phase so as to improve the semantic matching ability of the pre-trained model in downstream IR tasks. Different from them, we directly model the relationship between two hyperlinked documents. The rich semantic information contained in the documents can be better captured, which can facilitate the performance on downstream tasks. We present more detailed comparison between our method and existing studies in Section~\ref{sec:method3}.

\section{Methodology}

In this work, we propose a \textbf{P}rogressive \textbf{H}yperlink \textbf{P}rediction (PHP) task for pre-training. This task is devised based on our observation on human learning process. When faced with a new concept in a text, we often check its context to capture the background information. If such information is insufficient, we may refer to extra knowledge from other sources. The former step can be simulated by a masked language modeling task, which has already been applied in existing pre-trained models (\eg, BERT). The latter \textit{reference} step has not been fully explored, which is our focus.

In our proposed PHP task, we simulate the reference process by hyperlink. More specifically, we treat the anchor texts as the key concepts of a document. In addition to the context information, the hyperlinked documents also provide knowledge for understanding the anchor text. Different from existing studies that merely distinguish link/unlink relationship, we consider four kinds of link relationships with different strengths, which can reflect the documents' relevance. 
% In section \ref{sec:method1}, we introduce how we define these four kinds of relationships. In section \ref{sec:method2}, we introduce our pre-training strategy. In section \ref{sec:method3}, we compare PHP with previous works and illustrate why we can achieve state-of-the-art performance.

\subsection{Hyperlink Relationships} \label{sec:method1}

Given a document $d$ and a segment $s$ in $d$, we can divide all documents hyperlinked by $d$ (which we denote $\mc{D} = \{d_i\}$) into four groups based on the type of hyperlinks and the position where the hyperlinks occur. As we will see below, these four groups of documents have varying degrees of correlation with $s$. Before we look at them in detail, let's introduce a few notations. For a hyperlinked document $d_i$, let $a_i$ denote $d_i$'s corresponding anchor text in $d$, $a$ denote $d$'s corresponding anchor text in $d_i$ (if exists), and $\operatorname{Seg}(a, d_i)$ denote the id of the segment in $d_i$ where anchor text $a$ appears. We then categorize $\mc{D}$ into four subsets:

% Given a document $d$ and a segment $s$ in $d$ as our current context, we build a graph based on hyperlinks. Let $\mc{D} = \{d_i\}$ denotes the wiki document set made up of documents linked by document $d$, $a_i$ denotes $d_i$'s corresponding anchor text in $d$, $a$ denote $d$'s corresponding anchor text in $d_i$, and $\operatorname{Seg}(a, d_i)$ denote the id of the segment in $d_i$ where anchor text $a$ appears. We then categorize $\mc{D}$ into four subsets according to the relation between $d$ and $d_i$:
% and web pages in the set may link to each other. For a web page P in $S$, some web pages link to P and some web pages are linked by P. We can divide all these web pages except P into four groups according to their relation with P.

\begin{itemize}
    \item $\mc{D}_1$ - Strong symmetric linked documents in segment $s$ (such as Doc 1 in Figure \ref{fig:overview}): There exists a bidirectional hyperlink between $d$ and $d_i$. What's more, $a_i$ appears in the context $s$ and $a$ appears in the first segment of $d_i$. \ie, $d\leftrightarrow d_i \; \land \; a_i \in s \; \land \; \operatorname{Seg}(a, d_i) = 1$.
    \item $\mc{D}_2$ - Weak symmetric linked documents in segment $s$ (such as Doc 2 in Figure \ref{fig:overview}): There exists a bidirectional hyperlink between $d$ and $d_i$ and $a_i$ appears in the context $s$. Different from $\mc{D}_1$, $a$ here appears in a segment of $d_i$ other than the first. \ie, $d\leftrightarrow d_i \; \land \; a_i \in s \; \land \; \operatorname{Seg}(a, d_i) > 1$.
    \item $\mc{D}_3$ - Asymmetric linked documents in segment $s$ (such as Doc 3 in Figure \ref{fig:overview}): There exists a unidirectional hyperlink from $d$ to $d_i$ and $a_i$ indeed appears in the context $s$. \ie, $d \rightarrow d_i\;{\land}\;a_i\in s$;
    \item $\mc{D}_4$ - Asymmetric linked documents not in segment $s$ (such as Doc 4 in Figure \ref{fig:overview}): There exists a unidirectional hyperlink from $d$ to $d_i$ but $a_i$ does not appears in the segment $s$. \ie, $d \rightarrow d_i\;\land a_i \not\in s$;
\end{itemize}

Note that 
\begin{align*}
\mc{D}_1 \cup \mc{D}_2 \cup \mc{D}_3 \cup \mc{D}_4 = \mc{D},\;\;\mc{D}_i \cap \mc{D}_j = \varnothing \;\; if\;\; i \neq j.
\end{align*}

Intuitively, the former groups of web documents are more relevant to $s$ than the latter ones. Firstly, the first segment is often the description or summary of a wiki document, therefore there is a higher probability that hyperlinks in the first segment is more important than those in other segments. Depending on this observation, we categorize symmetric linked documents in $s$ into $\mc{D}_1$ and $\mc{D}_2$ and believe $\mc{D}_1$ is more relevant to $s$ than $\mc{D}_2$. Secondly, two documents linked to each other means they complement each other semantically, so the correlation is stronger than a asymmetric link. This provides a basis why documents in $\mc{D}_1$ and $\mc{D}_2$ are more related to $s$ than those in $\mc{D}_3$. Finally, documents in $\mc{D}_1$, $\mc{D}_2$ and $\mc{D}_3$ are hyperlinked from segment $s$ while those in $\mc{D}_4$ are hyperlinked from some segment in $d$ except $s$. This means documents in $\mc{D}_4$ would have relatively lower semantic correlation with $s$.

% Intuitively, the former groups of web documents are more relevant to $s$ than the latter ones. First, both documents in $\mc{D}_1$ and $\mc{D}_2$ are bidirectionally hyperlinked to $d$, however, for that the first segment is often the description or summary of the wiki document, linking back to $d$ from segment 1 means a stronger relevance to $d$. Similarly, comparing to unidirectional hyperlinks between $\mc{D}_3$ and $d$, bidirectional hyperlinks in $\mc{D}_2$ means that $d$ is also import to $d_i$, indicating a stronger relation. Finally, the difference between documents in $\mc{D}_3$ and $\mc{D}_4$ is that the anchor texts of documents in $\mc{D}_3$ appear in context $s$ while $\mc{D}_4$ does not. For that we use $s$ as our query, documents in $\mc{D}_3$ is obviously more relevant to $s$ than documents in $\mc{D}_4$.

\subsection{Pre-training Strategy} \label{sec:method2}
In this section, we introduce our pre-training objective based on hyperlink relationships introduced above and a supplementary masking strategy for anchor texts. The overview of our method is shown in Figure~\ref{fig:overview}.
% show our cascade pre-training strategy based on relationship differences. We will demonstrate how will we utilize

\subsubsection{Self-supervision Signal} As we discussed in \ref{sec:method1}, documents in former groups are more relevant to $s$ than the latter ones. As a pre-training object which resembles re-ranking task, we view segment $s$ as user query, which we denote $q$, and view documents in $\mc{D}$ as retrieved documents. Our query $q$ and documents $\mc{D}$ conform to the one-to-many pattern in re-ranking scenario. Formally, the relation between $q$ and documents in $\mc{D}$ can be demonstrated as:
\begin{align*}
\forall d_i \in \mc{D}_i, d_j \in \mc{D}_j,\quad S(q, d_i) > S(q, d_j) \;\; if \;\; i < j,
\end{align*}
where $S(q, d)$ is the relevance score between $q$ and $d$.

Our pre-training strategy is obtained by sampling documents from these four groups, and differences in their relevance to $q$ are used as the supervision signal of the pre-training objective.f

\subsubsection{Three-stage Pre-training Objectives} On the one hand, in all three stages, we use segment $s$ as our query $q$. On the other hand, in different stages, we take positive and negative samples from different groups of documents, resulting in three pre-training objectives ranging from easy to difficult. Concretely, 
\begin{itemize}
    \item Stage 1 - \textbf{H}yperlink \textbf{P}rediction (HP): Take positive samples from $\mc{D}_1 \cup \mc{D}_2 \cup \mc{D}_3$ and negative samples from $\mc{D}_4$.
    \item Stage 2 - \textbf{S}ymmetric \textbf{H}yperlink \textbf{P}rediction (SHP): Take positive samples from $\mc{D}_1 \cup \mc{D}_2$ and negative samples from $\mc{D}_3$.
    \item Stage 3 - \textbf{M}ost \textbf{R}elevant \textbf{D}ocument \textbf{S}election (MRDS): Take positive samples from $\mc{D}_1$ and negative samples from $\mc{D}_2$.
\end{itemize}

Therefore, in each stage, given the query $q$ we will get a positive document set $P(q)$ and a negative document set $N(q)$. For each query-document pair we concatenate them and use a Transformer structure to compute the relevance score:
\begin{align}
X_i &= \text{[CLS]}q\text{[SEP]}d_i\text{[SEP]}, \label{input_of_model} \\
S(q, d_i) &= \operatorname{MLP}(\operatorname{Transformer}_{\text{[CLS]}}(X_i)),  \label{prob_conditioned_on_d}
\end{align}
where $d_i$ comes from $P(q)$ or $N(q)$, the representation of ``[CLS]'' token is used to represent the sequence, and MLP($\cdot$) is a multi-layer perceptron.\footnote{The transformer model has the same structure as {BERT}~\cite{DBLP:conf/naacl/DevlinCLT19}}

We then use Localized Contrastive Estimation (LCE) loss to optimize the model, which is defined as:
\begin{align}
\mathcal{L}_{\text{PHP}}
&= - \sum_{d_i \in P(q)} \operatorname{log}\frac{\operatorname{exp}\left( S(q, d_i) \right)}{\operatorname{exp}\left( S(q, d_i) \right) + \sum_{d_j \in N(q)}\operatorname{exp}\left( S(q, d_j) \right)},
\end{align}
where from stage 1 to stage 3 we replace the notation PHP with HP, SHP and MRDS respectively.

\subsubsection{Mask Strategy for Anchor Text.} 

Our another training objective is Masked Language Modeling task which enables language models to learn good contextual representations. In our method, it helps to encode the detailed explanation of keywords into current context, which enriches the information of word representation.

The MLM loss is defined as following:
\begin{align}
\mathcal{L}_{\text{MLM}} 
&= \sum_{i} \sum_{j}\log p(x_{ij}|\hat{X}_i),
\end{align}
where $x_{ij}$ is the j-th token of $X_i$ and $\hat{X}_i$ is the masked edition of $X_i$.

In our method, the anchor text acts as a bridge between query and the corresponding document. To utilize the anchor text information, a mask based strategy is proposed. In the constructed document pair $(q, d_i)$, we sample the anchor text $a_i$ that $q$ links to $d_i$ with a probability of 50 percent and 15 percent for the left tokens. For the chosen tokens, the masking strategy is same to conventional bert mask.

It is \citet{DBLP:journals/corr/abs-2111-06377} claimed that languages are highly semantic and information-dense. Therefore, a high mask ratio for texts will hurt the model performance for that important information may lose. However, anchor texts are not only import information, but also closely related to the overall context and can be inferred by combining information from query and document. By increasing the mask portion of anchor texts to a specific ratio, such as 50\%, we can improve the model's ability to capture key information without losing too much important information.

\subsection{Discussion} \label{sec:method3}

\subsubsection{Utilizing Hyperlinks} There has been some pre-trained language models utilizing hyperlinks. However, they merely distinguish whether hyperlinks exist, rather than analyzing hyperlink relationships in depth. On the one hand, introducing linked documents into current context provides rich and closely related background knowledge for pre-training models to better understand texts and learn knowledge. On the other hand, the detailed sub-division of link relations encourages pre-training models to learn to distinguish documents of different relevance. Table \ref{tab:utilizing_hyperlinks} compares how pre-trained models utilize hyperlink relationships.

% Table generated by Excel2LaTeX from sheet '模型对比'
\begin{table}[t]
  \centering
  \small
  \caption{Comparison between pre-training methods in terms of how to utilizing hyperlinks.}
    \begin{tabular}{lccc}
    \toprule
          & \multicolumn{1}{c}{inside doc} & \multicolumn{1}{c}{cross doc} & \multicolumn{1}{c}{link strength} \\
    \midrule
    BERT  & \yes     & \noo     & \noo \\
    Transformer-ICT & \yes     & \noo     & \noo \\
    Transformer-WLP & \yes     & \yes     & \noo \\
    PROP(Wiki) & \yes     & \noo     & \noo \\
    HARP  & \yes     & \yes     & \noo \\
    LinkBERT & \yes     & \yes     & \noo \\
    PHP(Our) & \yes     & \yes     & \yes \\
    \bottomrule
    \end{tabular}%
  \label{tab:utilizing_hyperlinks}%
\end{table}%

\subsubsection{Pre-training for IR}Previous works such as WLP, ICT, PROP and HARP are designed based on the hypothesis that the pre-training objective which more closely resembles the target task leads to better fine-tuning performance. Therefore they construct pseudo query-document pairs in order to simulate the query-document matching problem in information retrieval scenario. However, some of them have ignored some import feature faced with information retrieval scenario, especially in re-ranking scenario. We summarize three key points to measure how similar these methods are to information retrieval task and the detailed comparison is illustrated in Table \ref{tab:pretrin_for_ir}:

% \begin{itemize}
    $\bullet$ Emphasize key information in queries. Queries in real word scenario are often composed of several keywords which may carry important information to understand user intention. PROP samples representative word sets from given documents while HARP builds pseudo query using anchor's corresponding sentence. We find that anchor text are a bridge connecting segment $s$ and the linked document, and propose a specially designed mask strategy for anchor text.
    
    $\bullet$  One query to many documents pattern. One important feature in information task is that given a query $q$, a huge number of candidate documents are waiting to be evaluated for their relevance score with $q$. WLP and ICT take negative samples from the mini-batch query-document pairs. However, PROP and HARP propose to simulate user queries by sampling word sets from a given document, resulting in many-queries-to-one-document pa ttern, which is totally different from the one-query-to-many-documents pattern in information retrieval. 
    
    $\bullet$  Strong negative samples. Models in IR are supposed to pick out most relevant documents to query $q$, especially in re-ranking scenario. Negative samples sampled from mini-batch in WLP and ICT can help models to distinguish relevant documents from irrelevant documents, but do not help further distinguish between related documents. In our proposed PHP, negative samples in all the three stages are linked documents, meaning they are all relevant to query $q$ in some degree. More importantly, as a progressive training strategy, negative samples in latter pre-training stages are more relevant to $q$ than in former ones, 
   helping models gradually gain a better ability to distinguish between related documents.
% \end{itemize}

% Table generated by Excel2LaTeX from sheet '模型对比'
\begin{table}[t]
  \centering
  \small
  \caption{Comparison between pre-training methods in terms of similarity to information retrieval task.}
    \begin{tabular}{lccc}
    \toprule
          & key info & one-to-many & strong neg. \\
    \midrule
    BERT  & \noo   & \noo   & \noo \\
    Transformer-ICT & \noo   & \yes   & \noo \\
    Transformer-WLP & \noo   & \yes   & \noo \\
    PROP(Wiki) & \yes   & \noo   & \yes \\
    HARP  & \yes   & \noo   & \yes \\
    LinkBERT & \noo   & \noo   & \noo \\
    PHP(Our) & \yes   & \yes   & \yes \\
    \bottomrule
    \end{tabular}%
  \label{tab:pretrin_for_ir}%
\end{table}%

\section{Experiments}
\subsection{Datasets and Evaluation Metrics}
\subsubsection{Pre-training Datasets}
We use English Wikipedia for pre-training.\footnote{\url{https://dumps.wikimedia.org/enwiki/}, enwiki dump progress on 20210101.} It contains tens of millions of documents with hyperlinks, which is suitable for our pre-training. Following previous work~\cite{DBLP:conf/coling/SunSQGHHZ20,DBLP:conf/wsdm/MaGZFJC21,DBLP:conf/cikm/MaDXZJCW21}, we use the public WikiExtractor to preprocess the download Wikipedia dump while preserving the hyperlinks.\footnote{\url{https://github.com/attardi/wikiextractor}} For data cleaning, we remove the documents with less than 100 words, resulting in 15,492,885 documents for pre-training. 

\subsubsection{Downstream Datasets} To validate the effectiveness of our proposed pre-training method, we fine-tune and evaluate our model on the ad-hoc retrieval task..
% , and compare its performance with the baselines.
Given an input query, the model is asked to rerank a list of candidate documents according to their relevance to the query. We use two datasets: 

(1) \textbf{{MS MARCO Document Ranking}}~\cite{DBLP:conf/nips/NguyenRSGTMD16} is a large-scale benchmark dataset for document retrieval task. It consists of 3.2 million documents with 367,000 training queries, 5,000 development queries, and 5,000 test queries. Binary labels are provided for indicating whether a document is relevant or not. 

(2) \textbf{{TREC 2019 Deep Learning (DL) Track}}~\cite{DBLP:journals/corr/abs-2003-07820} annotates 43 queries of MS MARCO as test queries. The relevance is scored in 0/1/2/3.

\subsubsection{Evaluation Metrics} For the ad-hoc retrieval task, we follo the official instructions~\cite{DBLP:conf/nips/NguyenRSGTMD16,DBLP:journals/corr/abs-2003-07820}, and employ MRR@\{10,100\} and nDCG@\{10,100\} to evaluate the ranking performance on MS MARCO and TREC DL, respectively. 
% As for the extractive QA task, we use F1 to measure the performance.

\subsection{Baselines}
We compare our method with several baselines, which can be categorized into three groups:

(1) \textbf{Traditional IR methods} are usually based on statistical models without using neural networks. 

$\bullet$~{QL} (Query likelihood)~\cite{DBLP:conf/sigir/ZhaiL01} is a language model with Dirichlet smoothing for ad-hoc retrieval. 

$\bullet$~{BM25}~\cite{DBLP:conf/sigir/RobertsonW94} is a classic ranking function that computes the relevance score based on the query terms appearing in each document.

(2) \textbf{Neural IR models} apply neural networks to computing relevance scores and ranking documents.

$\bullet$~{DRMM}~\cite{DBLP:conf/cikm/GuoFAC16} is a deep relevance matching model for ad-hoc retrieval. It employs a joint deep architecture at the query term level over the local interactions between query and document terms for relevance matching.

$\bullet$~{DUET}~\cite{DBLP:conf/www/Mitra0C17} consists of two separate deep neural networks that match the query and the document based on a local representation and learned distributed representations,  respectively. The two networks are jointly trained.

$\bullet$~{KNRM}~\cite{DBLP:conf/sigir/XiongDCLP17} uses a translation matrix to model word-level similarities, a kernel-pooling technique to extract multi-level matching features, and a learning-to-rank
layer that combines those features to compute the ranking score.

$\bullet$~{Conv-KNRM}~\cite{DBLP:conf/wsdm/DaiXC018} is an extension of {KNRM}. It apples Convolutional Neural Networks to represent $n$-grams of various lengths and constructs translation matrices. The following computation is the same as {KNRM}.

All the aforementioned models are directly trained on the downstream tasks without pre-training.

(3) \textbf{Pre-trained models} are deep neural networks pre-trained on large-scale corpus. They are usually fine-tuned on IR datasets for retrieval tasks. 

$\bullet$~{BERT}~\cite{DBLP:conf/naacl/DevlinCLT19} is a transformer-based model pre-trained with Masked Language Modeling (MLM) and Next Sentence Prediction (NSP) objectives. It has become a strong baseline model for ad-hoc retrieval due to its powerful contextual language representations. We use the pre-trained model provided by Huggingface and fine-tune it on the donwstream datasets.\footnote{\url{https://huggingface.co/bert-base-uncased}}

$\bullet$~{Transformer+ICT}~\cite{DBLP:conf/iclr/ChangYCYK20} follows the architecture and hyperparameters of {BERT}, but trains the Transformer model with Inverse Cloze Task (ICT) and MLM from scratch. ICT is designed specifically for passage retrieval, which aims at predicting a removed sentence based on its context. 

$\bullet$~{Transformer+WLP}~\cite{DBLP:conf/iclr/ChangYCYK20} is similar to the former method, but trains the Transformer model with Wiki Link Prediction (WLP) and MLM. Considering a random sentence in the first paragraph of a Wikipedia page, WLP aims to predict the passage that is linked to the sentence via a hyperlink. 
We pre-train the former two Transformer models on Wikipedia corpus and fine-tune them on downstream datasets for a fair comparison.

$\bullet$~{PROP}~\cite{DBLP:conf/wsdm/MaGZFJC21} pre-trains BERT with Representative wOrds Prediction (PROP) and MLM for ad-hoc retrieval. The PROP objective trains the model to predict the pairwise preference between two words sets. We experiment with the released models pre-trained on Wikipedia, which is denoted as {PROP (Wiki)}. 

$\bullet$~{HARP}~\cite{DBLP:conf/cikm/MaDXZJCW21} pre-trains BERT with four self-supervised training objectives and MLM. It also leverages hyperlinks by using anchor texts as pseudo queries and learning their relationship with linked documents. This is the state-of-the-art pre-training method for IR.

\begin{table}[t]
    \centering
    \small
    \caption{Experimental results of all models on two ad-hoc retrieval datasets. The best results are in bold. $\dag$ indicates our {PHP} achieves significant improvement in t-test with $p$-value $<$ 0.05.}
    {
    \setlength{\tabcolsep}{1.0mm}
    \begin{tabular}{lcccc}
    \toprule
        & \multicolumn{2}{c}{ANCE Top 100} & \multicolumn{2}{c}{Official Top 100} \\
        % \cmidrule(lr){2-3}\cmidrule(lr){4-5}
        \textbf{{MS MARCO}} & MRR@100 & MRR@10 & MRR@100 & MRR@10 \\
        \midrule
        {QL} & .2457 & .2295 & .2103 & .1977 \\
        {BM25} & .2538 & .2383 & .2379 & .2260 \\
        \midrule
        {DRMM} & .1146 & .0943 & .1211 & .1047 \\
        {DUET} & .2287 & .2102 & .1445 & .1278 \\
        {KNRM} & .2816 & .2740 & .2128 & .1992 \\
        {Conv-KNRM} & .3182 & .3054 & .2850 & .2744 \\
        \midrule
        {BERT} & .4184 & .4091 & .3830 & .3770 \\
        {Transformer-ICT} & .4194 & .4101 & .3828 & .3767 \\
        {Transformer-WLP} & .3998 & .3900 & .3698 & .3635 \\
        {PROP (Wiki)} & .4188 & .4092 & .3818 & .3759 \\
        {HARP} & .4472 & .4393 & {.4012} & {.3961} \\
        % {LinkBERT} & .4522 & .4446 & .4130 & .4085 \\
        {PHP (Our)} & \textbf{.4601}$^\dag$ & \textbf{.4526}$^\dag$ & \textbf{.4155}$^\dag$ & \textbf{.4106}$^\dag$ \\
        \midrule
        \midrule
        & \multicolumn{2}{c}{ANCE Top 100} & \multicolumn{2}{c}{Official Top 100} \\
        % \cmidrule(lr){2-3}\cmidrule(lr){4-5}
        \textbf{{TREC DL}} & nDCG@100 & nDCG@10 & nDCG@100 & nDCG@10 \\
        \midrule
        {QL} & .4644 & .5370 & .4694 & .4354 \\
        {BM25} & .4692 & .5411 & .4819 & .4681 \\
        \midrule
        {DRMM} & .3812 & .3085 & .4099 & .3000 \\
        {DUET} & .3912 & .3595 & .4213 & .3432 \\
        {KNRM} & .4671 & .5491 & .4727 & .4391 \\
        {Conv-KNRM} & .4876 & .5990 & .5221 & .5899 \\
        \midrule
        {BERT} & .4900 & {.6084} & .5289 & .6358 \\
        {Transformer-ICT} & {.4906} & {.6053} & .5300 & {.6386} \\
        {Transformer-WLP} & {.4891} & .6143 & .5245 & .6276 \\
        {PROP (Wiki)} & .4882 & .6050 & .5251 & .6224 \\
        {HARP} & {.4949} & .6202 & {.5337} & .6562 \\
        % {LinkBERT} & .5004 & .6335 & .5374 & .6563 \\
        {PHP (Our)} & \textbf{.5045}$^\dag$ & \textbf{.6481}$^\dag$ & \textbf{.5401} & \textbf{.6612} \\
    \bottomrule
    \end{tabular}
    }
    \label{tab:main_experiment}
\end{table}

% Table generated by Excel2LaTeX from sheet 'Sheet1'
\begin{table*}[t]
  \centering
  \small
  \caption{Performance (F1) on six extractive question answering datasets. The best results are in bold. $\dag$ indicates our {PHP} achieves significant improvement in t-test with $p$-value $<$ 0.05.}
    \begin{tabular}{lccccccc}
    \toprule
          & HotpotQA & TriviaQA & SearchQA & NaturalQ & NewsQA & SQuAD & Avg. \\
    \midrule
    BERT  & 75.71  & 68.63  & 75.37  & 77.81  & 65.25  & 87.62  & 75.06  \\
    Transformer-ICT & 75.02  & 69.07  & 75.25  & 77.82  & 66.82  & 87.88  & 75.31  \\
    Transformer-WLP & 73.10  & 67.28  & 74.78  & 76.67  & 65.82  & 87.22  & 74.14  \\
    PROP(Wiki) & 75.29  & 69.14  & 75.98  & 77.85  & 65.45  & 88.30  & 75.34  \\
    HARP  & 76.04  & 69.01  & 76.09  & 77.96  & 65.68  & 87.97  & 75.46  \\
    % LinkBERT & 77.77  & 72.84  & 77.87  & 78.16  & 68.45  & \textbf{90.37 } & 77.58  \\
    PHP (Our) & \textbf{77.93}$^\dag$ & \textbf{75.20}$^\dag$ & \textbf{78.24}$^\dag$ & \textbf{79.50}$^\dag$ & \textbf{69.46}$^\dag$ & \textbf{89.96}$^\dag$ & \textbf{78.38}$^\dag$ \\
    \bottomrule
    \end{tabular}%
  \label{tab:qa}%
\end{table*}%

% \begin{table*}[t]
%   \centering
%   \small
%   \caption{Performance on the GLUE benchmark. All tasks are evaluated on the development set. The best results are in bold.}
%     \begin{tabular}{lccccccccc}
%     \toprule
%           & MNLI-(m/mm) & QQP   & QNLI  & SST-2 & CoLA  & STS-B & MRPC  & RTE   &  Avg. \\
%     \midrule
%     BERT  & 84.31/84.95 & 90.88 & 91.07 & \textbf{93.00} & 56.76 & 88.53 & 88.78 & 66.43 & 82.51 \\
%     Transformer-ICT & 84.08/83.97 & 90.67 & 90.55 & 91.28 & 54.03 & 88.27 & 85.91 & 61.73 & 80.81 \\
%     Transformer-WLP & 84.54/84.81 & 91.03 & 91.36 & 92.43 & 57.01 & 88.89 & 88.22 & 65.34 & 82.37 \\
%     PROP(Wiki) & 84.36/84.81 & 90.97 & 91.21 & 92.66 & 52.60 & 88.30 & 87.87 & 68.59 & 82.10 \\
%     HARP  & 84.37/84.83 & 90.88 & 91.27 & 92.20 & \textbf{57.78} & 88.98 & 87.50 & 69.68 & 82.86 \\
%     LinkBERT & 84.66/84.22 & 90.73 & 91.84 & 92.78 & 53.64 & 88.35 & 89.35 & 70.76 & 82.80 \\
%     PHP (Our) & \textbf{84.85/84.96} & \textbf{91.06} & \textbf{92.17} & 92.43 & 56.78 & \textbf{88.99} & \textbf{90.78} & \textbf{75.09} & \textbf{84.02} \\
%     \bottomrule
%     \end{tabular}%
%     \vspace{-8px}
%   \label{tab:glue}%
% \end{table*}%

\subsection{Implementation Details}

\subsubsection{Model Architecture}
For our method {PHP}, we use the same Transformer encoder architecture as the base {BERT}~\cite{DBLP:conf/naacl/DevlinCLT19}. The hidden size is 768, and the number of self-attention heads is 12. 
% For a fair comparison, all of the pre-trained baseline models use the same architecture as our model. 
We use the HuggingFace's Transformers for the model implementation~\cite{2019HuggingFace}.

\subsubsection{Pre-training Settings}
% During the pre-training process, we adopt the Reranker framework proposed by \citet{DBLP:conf/ecir/GaoDC21} which use Localized Contrastive Estimation(LCE) for training.
For the MLM objective, we randomly select 50\% words for anchor texts and 15\% words for other tokens. Then, following the operation in vanilla {BERT}, for all selected tokens, 80\% of them are replaced with the [MASK] token, 10\% are replaced with a random token, and 10\% remain unchanged. We use the Adam optimizer with a learning rate of 1e-5, a warm up ratio of 0.1, and a weight decay of 0.01. The model is trained for one epoch with hyperlink prediction, one epoch with symmetric hyperlink prediction, and two epochs with most relevant document prediction. During all these three stages, the batch size is set as 24 and the maximum length of the input sequence is set as 512. 
% We use $\text{BERT}_{\text{base}}$ to initialize our model. 
For our multi-stage pre-training, models in later stages inherit the parameters of the model in earlier stages.

\subsubsection{Fine-tuning Settings}
% During fine-tuning, we also adopt LCE framework. 
% The learned parameters after pre-training are used to initialize the fine-tuning model. 
We follow the previous work~\cite{DBLP:conf/cikm/MaDXZJCW21, DBLP:conf/wsdm/MaGZFJC21, DBLP:journals/corr/abs-1901-04085} and test the performance of all methods on the re-ranking task. To test the performance with different quality of the candidate document set, we re-rank the documents from two candidate sets, $\ie$, ANCE Top100 and Official Top100. ANCE Top100 is retrieved based on the ANCE model~\cite{DBLP:conf/iclr/XiongXLTLBAO21}, and Official Top100 is released by the official MS MARCO and TREC teams. During fine-tuning, we concatenate the URL, title, and body of one document as the document content and concatenate the query and document as the input. The batch size is set as 32, and the maximum length of the input sequence is also 512. We fine-tune for one epoch using Adam optimizer, with a learning rate of 1e-5, a warm-up ratio of 0.01, and a weight decay of 0.1.

\subsection{Experimental Results}
For MS MARCO, we evaluate all methods on the development set. For TREC DL, we report the performance on the test set. The results are shown in Table~\ref{tab:main_experiment}. In general, we can see {PHP} outperforms all baseline methods in terms of all evaluation metrics. This result clearly demonstrates the superiority of our method. Based on the results, we also have the following observations. 

(1) Traditional models such as {QL} and {BM25} are strong baselines on both datasets, but they cannot capture semantic relevance. Instead, neural IR models apply deep neural networks to measure the semantic relevance based on distributed representation of query and documents. Therefore, {KNRM} and {Conv-KNRM} can achieve better performance than traditional methods. Furthermore, pre-trained models significantly outperform other methods. This indicates their great potential on modeling relevance between query and documents. The advantage may stem from two perspectives: tremendous parameters in pre-trained models increase model capacity; and the general knowledge obtained from the pre-training process is beneficial for downstream tasks.

(2) Compared with {BERT}, models pre-trained with objectives tailored for IR can perform better. For example, {Transformer-ICT} achieves better performance on both datasets, demonstrating the effectiveness of the retrieval-relevant pre-trained objective. Intriguingly, {Transformer-WLP} performs worse than {BERT}. A potential reason is the noise introduced by WLP when constructing pseudo query-response pairs. 
% {PROP} is pre-trained with the ROP task, which aims at identifying representative words from the documents. 
In contrast, our {PHP}, {HARP}, and {LinkBERT} leverage hyperlink relationship to design pre-training objectives, and achieve better results. This confirms the potential of hyperlinks on modeling document relevance.

(3) {PHP} performs better than HARP. Specifically, {PHP} improves MRR@100 and MRR@10 by more than 3\% on the MS MARCO dataset with ANCE Top 100 results. {HARP} is the state-of-the-art pre-training method for ad-hoc IR, which also leverages hyperlinks. They focus on learning the relationship between anchor text and the linked document to simulate the relationship between query and documents. The improvement obtained by our method indicates that the potential of hyperlinks has not been fully explored by {HARP}, while our proposed strategy is much more effective. By progressively identifying hyperlinked documents in different groups, {PHP}'s capability of measuring relevance can be gradually enhanced.

% (3) \textbf{Our model SHP outperforms previous link-based models like {HARP}~\cite{DBLP:conf/cikm/MaDXZJCW21} and {Transformer-WLP}~\cite{DBLP:conf/iclr/ChangYCYK20} with a large margin.} {HARP} and {Transformer-WLP} are pre-trained models which util wiki link too. However, they limit their utilization on the existence of hyperlinks and do not further distinguish symmetric hyperlinks from asymmetric hyperlinks. The experiment results show that the potential of hyperlinks are far from fully explored. Concretely, SHP significantly outperforms {Transformer-WLP} by 6.39\% in terms of MRR@100 on MS MARCO ANCE Top100 and 7.39\% in terms of MRR@100 on MS MARCO Official Top 100. Besides, SHP outperforms {HARP} by 3.27\% in terms of MRR@100 on MS MARCO ANCE Top100 and 2.60\% in terms of MRR@100 on MS MARCO Official Top 100. On TREC DL Official Top 100, SHP outperforms {Transformer-WLP} and {HARP} by 4.74\% and 3.17\% respectively in terms of nDCG@10.

% (4) \textbf{Pre-trained methods outperform methods without pre-training and objectives tailored for IR can further increase performance.}

% \subsubsection{Results on Extractive QA}

\begin{table}[t!]
    \centering
    \small
    \caption{Ablation study of our method on the MS MARCO dataset.}
    \begin{tabular}{lcccc}
        \toprule
         & \multicolumn{2}{c}{ANCE Top 100} & \multicolumn{2}{c}{Official Top 100} \\
        %  \cmidrule(lr){2-3}\cmidrule(lr){4-5}
         & MRR@100 & MRR@10 & MRR@100 & MRR@10 \\
         \midrule
         {PHP (Full)} & \textbf{.4601} & \textbf{.4526} & \textbf{.4155} & \textbf{.4106} \\
         \quad {Only HP} & .4456 & .4377 & .4036 & .3987 \\
         \quad {Only SHP} & .4528 & .4447 & .4114 & .4066 \\
         \quad {Only MRDS} & .4513 & .4434 & .4107 & .4060 \\
        %  \quad {w/o HP} & \textbf{.4558} & \textbf{.4479} & .4132 & .4084 \\
        %  \quad {w/o Stage 3} & \\
        \bottomrule
    \end{tabular}
    \label{tab:ab}
\end{table}

\subsection{Further Analysis}
We further analyze our proposed {PHP} in several aspects as follows.
% To further analyze our proposed method, we investigate the following research questions (RQs):

% \begin{itemize}
%     \item RQ1: What is the influence of the three proposed pre-training objectives?
%     \item RQ2: How does {PHP} perform on few-shot and large-scale queries?
%     \item RQ3: How many negative samples are required in our pre-training objectives?
%     \item RQ4: What is the impact of anchor text mask ratio? 
%     \item RQ5: What is the influence of masked language modeling?
% \end{itemize}
\subsubsection{Ablation Study}
We conduct an ablation study to explore the influence of the three proposed pre-training objectives. Specifically, we pre-train the model with only one of the three pre-training objectives, and test the performance on the MS MARCO dataset. We denote the hyperlink prediction, symmetric hyperlink prediction, and most relevant document selection as {HP}, {SHP}, and {MRDS}, respectively. The experimental results are shown in Table~\ref{tab:ab}, and we can see:

First, pre-training with three progressive objectives performs best. This is consistent with our assumption that these objectives can gradually enhance the model's performance. Second, the symmetric hyperlink prediction task is the most effective. The potential reasons include: (1) Hyperlink prediction is much easier than symmetric hyperlink prediction, because the negative documents may be extremely irrelevant. As for the most relevant document selection, its performance may be limited by the scarce training samples (the negative documents should also be symmetric linked). As a result, among three pre-training tasks, symmetric hyperlink prediction can provide most sufficient knowledge, which is beneficial for downstream IR tasks. (2) Symmetric hyperlink prediction fit the downstream ad-hoc retrieval task most, because the candidate documents are relevant (given by either ANCE or official search engines), and the evaluation metrics focus on top-10/100 results rather than the top-1 performance. Finally, compared with {HARP}, pre-training with any one of our proposed objectives can improve the performance. This demonstrates once again the advantage of our method on utilizing hyperlinks. 

\subsubsection{Extension to Other Tasks}
We have evaluated our model on ad-hoc retrieval task. To further validates its generalizability, we fine-tune our method on both extractive question answering and natural language understanding tasks, which are widely used benchmarks. 

Extractive question answering task is formulated as: Given a set of documents and a question, this task aims to identify an answer span from the document. We consider six commonly used datasets from MRQA shared tasks~\cite{DBLP:conf/acl-mrqa/FischTJSCC19}: HotpotQA~\cite{DBLP:conf/emnlp/Yang0ZBCSM18}, TriviaQA~\cite{DBLP:conf/acl/JoshiCWZ17}, NaturalQ~\cite{DBLP:journals/tacl/KwiatkowskiPRCP19}, SearchQA~\cite{DBLP:journals/corr/DunnSHGCC17}, NewsQA~\cite{DBLP:conf/rep4nlp/TrischlerWYHSBS17}, and SQuAD~\cite{DBLP:conf/emnlp/RajpurkarZLL16}. As the MRQA shared task does not have a public test set, following previous work~\cite{DBLP:journals/corr/abs-2203-15827}, we split the dev set in half to make new dev and test sets. GLUE~\cite{DBLP:conf/iclr/WangSMHLB19} is a multi-task benchmark for natural language understanding. We use eight datasets from GLUE, and they are designed for several tasks, such as natural language inference, question answering, and text similarity. We follow the fine-tuning method BERT~\cite{DBLP:conf/naacl/DevlinCLT19} uses for these tasks. 

The experimental results are shown in Table~\ref{tab:qa} and Table~\ref{tab:glue}. Overall, we can see our PHP achieves the best results in terms of the average score.  Different from ad-hoc retrieval, the question answering task usually requires the model to reason over (multiple) documents. The better performance of PHP indicates that our utilization of hyperlinks are more effective. Besides, we also observe improvement over several natural language understanding tasks, this validates our assumption that using additional knowledge can facilitate the understanding of concepts.

\subsubsection{Low-resource and Large-scale Performance}
In actual applications, it is not only time-consuming but also expensive to collect large-scale and high-quality queries and documents for fine-tuning. Pre-trained models with a large number of parameters may suffer from the low-resource settings because small scale data often lead to model overfitting. Our method provides a training mode with progressive difficulty for the model based on hyperlinks, which means that the model can learn different information retrieval patterns with different difficulties and types during the pre-training phase.

\begin{figure}[t]
    \centering
    \includegraphics[width=.9\linewidth]{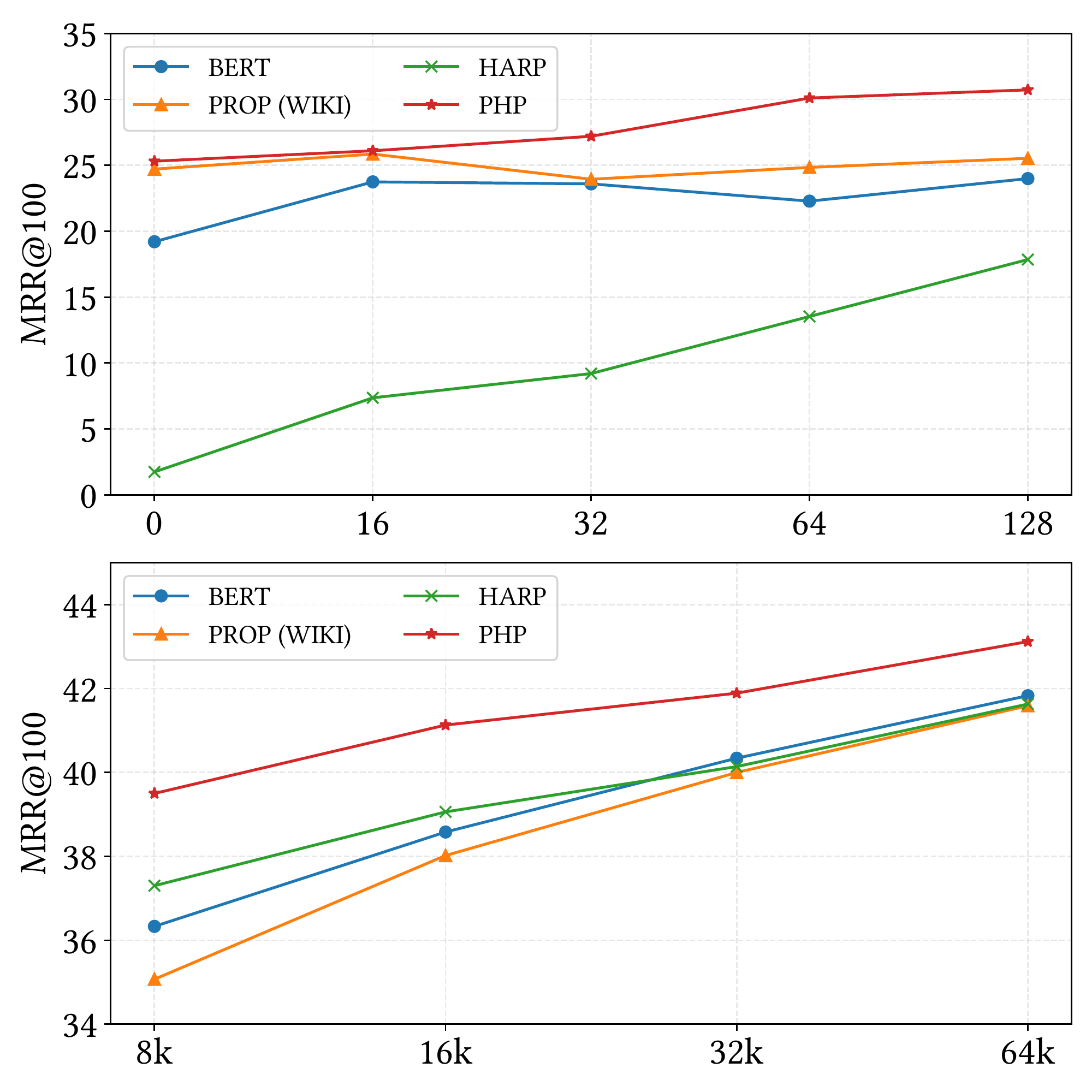}
    \caption{The evaluation results of few-shot (top) and large-scale (bottom) queries on the MS MARCO dataset with ANCE Top 100 documents.}
    \label{fig:few}
\end{figure}

In this section, we investigate the performance of our model {PHP} and baseline models on low-resource and large-scale settings respectively. For low-resource settings, we evaluate models on both zero-shot scenario, where there is no fine-tuning, and few-shot scenario, where we randomly pick 16/32/64/128 queries to fine-tune. For large-scale settings, we randomly pick 8k/16k/32k/64k queries. We compare the models in terms of MRR@100 on MS MARCO ANCE Top 100. The results are illustrated in Figure \ref{fig:few}.

(1) Under low-resource settings, our model {PHP} obtains a steady increase in performance and achieves the best results among all models when the number of queries is 64 and 128. Although {PROP-WIKI} wins the first place under zero-shot setting, its performance does not continue to increase when the number of queries increases, and is finally beaten by {PHP}. This means that our method has high robustness, both under low-resource settings and real-world practice, which is good performance for a kind of pre-trained models for information retrieval.\footnote{We notice the poor performance of {HARP} under low-resource settings. Since the authors do not release their pre-trained model, these results are obtained by our implementation based on their released code.}
%This means our pre-trained strategy tailored for IR tasks well under low-resource settings and can be applied in real-world practice.
(2) Under large-scale settings, our model shows great power and outperform baseline models with a large margin. Although the gap of performance between {PHP} and baseline models appears a trend of narrowing, the advantage keeps until the end of fine-tuning period as shown in Table\ref{tab:main_experiment}. In the case of sufficient fine-tuning data, our method can make full use of its advantages and exploit the potential of neural ranking models.
%This implies that {PHP} is indeed a excellent pre-training strategy that can exploit the potential of neural ranking model after being fully fine-tuned.

\subsubsection{Impact of Negative Samples}
\begin{figure}[t]
    \centering
    \includegraphics[width=\linewidth]{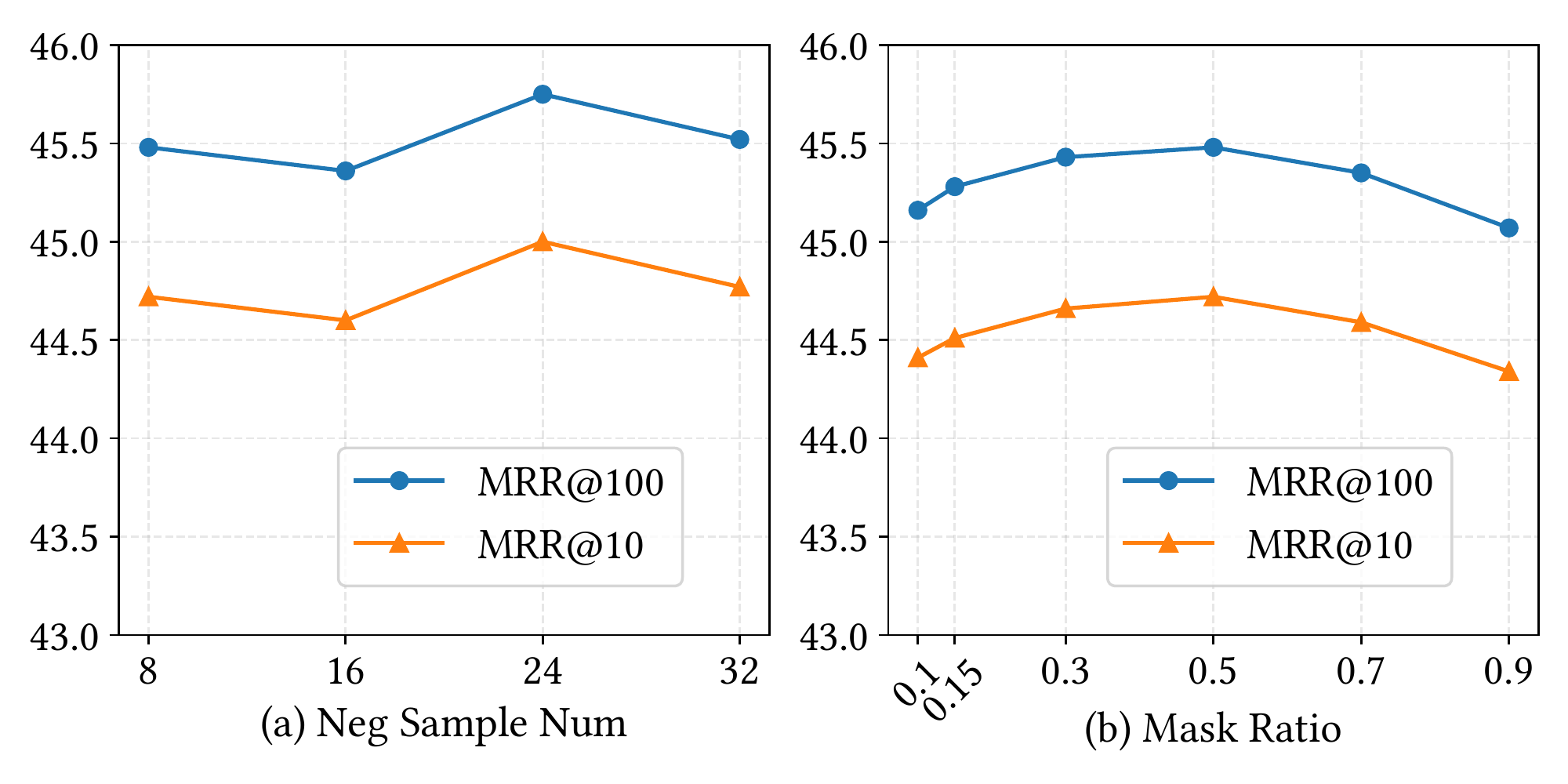}
    \caption{Performance of different numbers of negative samples (left) and different anchor text mask ratios (right). The results are on the MS MARCO validation set with ANCE Top 100 documents.}
    \label{fig:hyper}
\end{figure}

As a listwise loss function, the performance of LCE is influenced by the number of negative samples. To explore such influence, we conduct an experiment on the MS MARCO dataset by using different number of negative documents. For simplicity, in this experiment, we set the number of negatives as the same for the three pre-training tasks. The experimental results are illustrated in Figure~\ref{fig:hyper} (a). We can observe:

\textbf{First}, when the number of negatives is small ($<$ 24), the performance is worse. This implies that the negative documents are very important for modeling relevance. \textbf{Then}, when more negative documents used, they can provide more sufficient information for the model, so the performance is improved. {PHP} achieves the best performance when using 24 negative documents. \textbf{Finally}, {PHP}'s performance drops when 32 negative documents. We attribute this to the noise introduced by the sampling process. Since the number of negative documents is extremely limited in the most relevant document selection task, we apply the repeated sampling strategy. However, this may reduce the quality of the training samples. 

\subsubsection{Impact of Anchor Text Mask Ratio}
The performance of language models based on masked language modeling is very sensitive to the mask ratio of tokens. For most of the tokens, we follow BERT to randomly mask 15\% of them, while for anchor texts in pair $(d, d_i)$, we test the impact of different mask ratio in this section. We find that with the proportion of anchor text masking from low to high, the performance of the model first increases and then decreases. The experimental results are illustrated in Figure~\ref{fig:hyper} (b).

\textbf{First}, the performance curve takes on the shape of parabola and reaches the best performance when the mask ratio of anchor texts equals to 50\%. From the experimental results, we can see that the anchor text is crucial information connecting $d$ and $d_i$. $d$ is where the anchor text appears and provides the bridge information. For $d_i$, the anchor text may be the title or topic to some extend. With proper masking for them, the model tries it best to understand the semantic information in the pairs to fill them, which allows the model to obtain a more efficient semantic representation of input pairs. Increasing the mask ratio from 15\% to 50\% can force model to pay more attention to the anchor texts and therefore better encode the context, which improves the performance. However, as the ratio rises from 50\% to 90\%, the anchor text predicting task is getting harder, which makes the bridge information loses gradually. We find that 50\% is a balance point. \textbf{Second}, when the ratio changes from 50\% to 15\%, the decrease rate of model performance is lower than that when the ratio changes from 50\% to 90\%. This means that less mask proportion reduces the pre-trained difficulty, but the important information of anchor texts is not dropped too much. 
%We think the reason is that when the ratio decreases from 50\% to 10\%, the difficulty of the anchor text predicting task becomes lower, however, this does not hurt the model. 
On the other hand, when the ratio changes from 50\% to 90\%, the performance reduces rapidly. This side demonstrates that the information of anchor text is important. Small-scale masking is conducive to the enhancement of learning difficulty and improve robustness, and the large-scale masking will lose performance. \textbf{Finally}, mask ratio of 50\% is better than the default ratio of 15\%. Although the mask strategy of BERT works most of the time, it is still a good choice to try to change the mask ratio of certain words in certain tasks.

\begin{figure}[t]
    \centering
    \includegraphics[width=\linewidth]{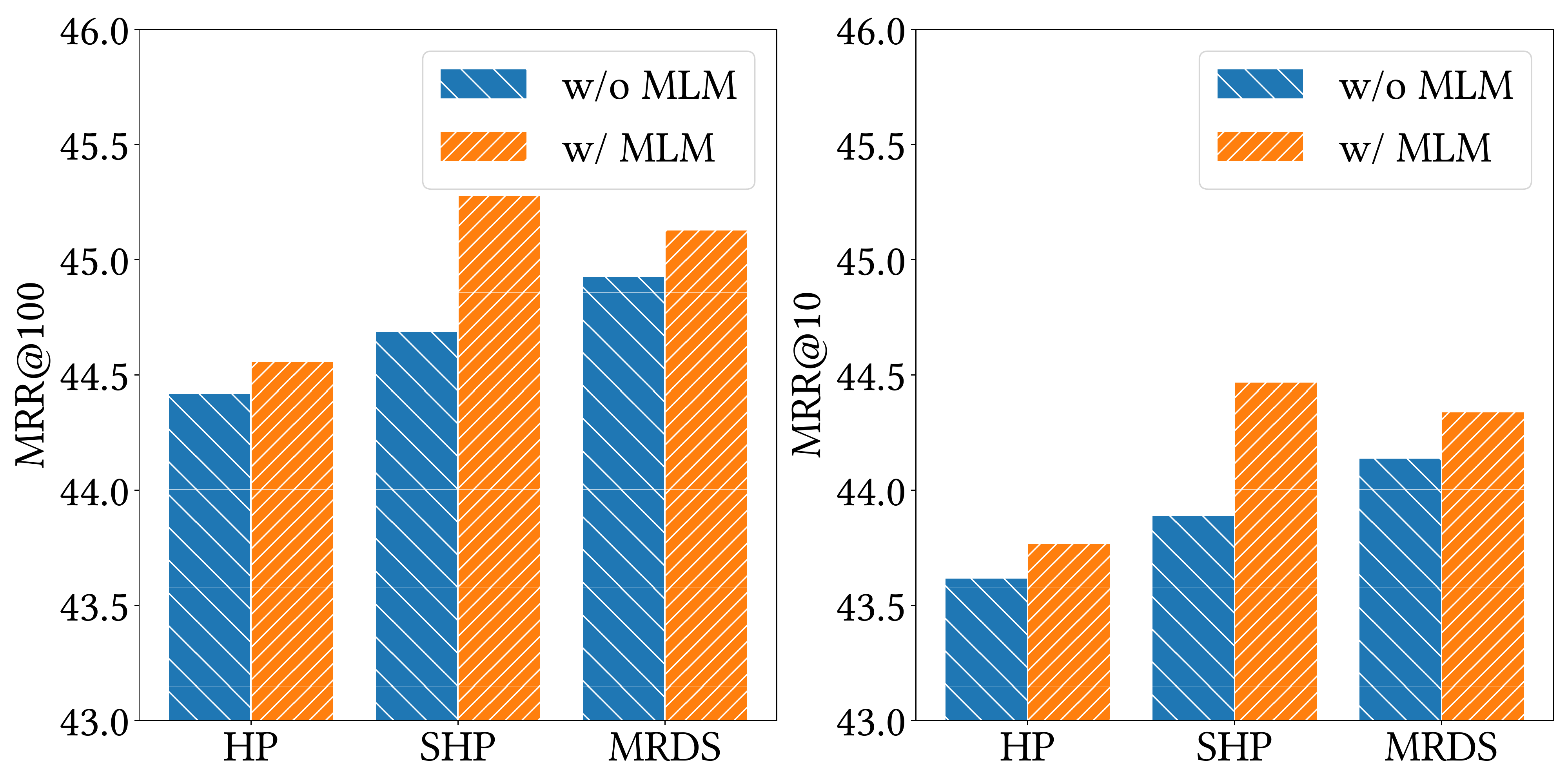}
    \caption{Performance of pre-training with or without masked language modeling. The results are on the MS MARCO dataset with ANCE Top 100 documents.}
    \label{fig:mlm}
\end{figure}

\subsubsection{Impact of Masked Language Modeling}
Masked language modeling (MLM) is a pre-training task proposed in {BERT}~\cite{DBLP:conf/naacl/DevlinCLT19}. It aims at predicting masked tokens through their context, which is very important for language modeling. In our {PHP}, we combine each of our proposed pre-training strategies with MLM. In this section, we perform a study to investigate the influence of MLM in our framework. The experimental results are shown in Figure~\ref{fig:mlm}, and we can observe:

\textbf{First}, MLM is an effective pre-training objective for {PHP}. Removing it from any proposed pre-training task leads to performance degradation. Indeed, our proposed pre-training tasks focus more on modeling the relationship across documents, which can be treated as high-level information. In contrast, MLM can provide a local view at token-level, which can be complementary with our hyperlink based tasks. \textbf{Second}, even pre-training without MLM, our {PHP} can still perform better than the state-of-the-art method {HARP} (with MLM). This confirms the improvement of our {PHP} stems from our proposed progressive pre-training tasks rather than merely involving MLM. 

\section{Conclusion and Future Work}
In this work, we proposed a progressive pre-training strategy for ad-hoc retrieval, which consists of three hyperlink-based pre-training tasks, namely hyperlink prediction, symmetric hyperlink prediction, and most relevant document selection. These pre-training tasks can simulate different stages of a retrieval process, and gradually enhance the model's performance on relevance modeling. Compared with existing methods, our strategy can make better use of utilize the hyperlink, which reveal the great potential of hyperlink in learning relevance modeling. The experimental results demonstrated the effectiveness of our model. In the future, we plan to integrate our method with other pre-training objectives (such as ROP), which may bring further improvement.

% % Table generated by Excel2LaTeX from sheet 'few shot'
% \begin{table}[htbp]
%   \centering
%   \caption{Experiment results of few-shot setting}
%     \begin{tabular}{lrrrrr}
%           & 0     & 16    & 32    & 64    & 128 \\
%     BERT  & 0.1920  & 0.2374  & 0.2359  & 0.2228  & 0.2399  \\
%     PROP-WIKI & 0.2471  & 0.2585  & 0.2394  & 0.2484  & 0.2553  \\
%     HARP  & 0.0174  & 0.0736  & 0.0920  & 0.1353  & 0.1785  \\
%     LINKBERT & 0.0922  & 0.1168  & 0.1674  & 0.2224  & 0.2846  \\
%     PHP   & 0.2561  & 0.2588  & 0.2713  & 0.2994  & 0.3110  \\
%     \end{tabular}%
%   \label{tab:addlabel}%
% \end{table}%

% % Table generated by Excel2LaTeX from sheet 'few shot'
% \begin{table}[htbp]
%   \centering
%   \caption{Experiment results of large-scale setting}
%     \begin{tabular}{lrrrr}
%           & \multicolumn{1}{l}{8k} & \multicolumn{1}{l}{16k} & \multicolumn{1}{l}{32k} & \multicolumn{1}{l}{64k} \\
%     BERT  & 0.3633  & 0.3858  & 0.4034  & 0.4183  \\
%     PROP-WIKI & 0.3507  & 0.3802  & 0.4000  & 0.4159  \\
%     HARP  & 0.3730  & 0.3906  & 0.4014  & 0.4163  \\
%     LINKBERT & 0.3803  & 0.3913  & 0.4072  & 0.4202  \\
%     PHP   & 0.3950  & 0.4113  & 0.4189  & 0.4312  \\
%     \end{tabular}%
%   \label{tab:addlabel}%
% \end{table}%

% \clearpage
%%% -*-BibTeX-*-
%%% Do NOT edit. File created by BibTeX with style
%%% ACM-Reference-Format-Journals [18-Jan-2012].

\end{document}